\documentclass[conference]{IEEEtran}
\IEEEoverridecommandlockouts

\usepackage{cite}
\usepackage{amsmath,amssymb,amsfonts}
\usepackage{algorithmic}
\usepackage{graphicx}
\usepackage{textcomp}
\usepackage{xcolor}
\usepackage{subcaption}
\usepackage{booktabs}
\usepackage{graphicx}
\usepackage{svg}
\usepackage{stfloats} 

\usepackage{xspace}
\usepackage{balance}
\usepackage{adjustbox}

\newcommand{\OurMethod}{\textsc{ChannelKAN}\xspace}

\def\BibTeX{{\rm B\kern-.05em{\sc i\kern-.025em b}\kern-.08em
    T\kern-.1667em\lower.7ex\hbox{E}\kern-.125emX}}

\begin{document}


\title{\OurMethod: Multi-Scale Dual-Domain Channel Prediction via Hybrid CNN-KAN Architecture}

\author{Nanqing Jiang,
Zhangyao Song,
Tao Guo,
Xiaoyu Zhao,
Yinfei Xu\\
\IEEEauthorblockA{School of Cyber Science and Engineering, Southeast University, Nanjing, China}
\IEEEauthorblockA{Email: {\tt \{nqjiang, zysong, taoguo, xy-zhao, yinfeixu\}@seu.edu.cn}}
}

\maketitle

\begin{abstract}
    Accurate channel state information (CSI) prediction is essential for improving the reliability and spectral efficiency of massive MIMO-OFDM systems in high-mobility scenarios.
    Existing deep learning methods struggle to jointly capture short-term local variations and long-range nonlinear dependencies in CSI sequences.
    To address this challenge, we propose \OurMethod, a hybrid CNN-KAN channel prediction model with multi-scale frequency domain information enhancement.
    The key insight is that CNNs and Kolmogorov-Arnold Networks (KANs) are naturally complementary: CNNs extract intra-time-step local spatial-frequency correlations, while KANs with learnable Chebyshev polynomial activations fit inter-time-step nonlinear temporal evolution in a holistic manner.
    Specifically, a dual-domain expansion module first generates complementary frequency-domain and delay-domain CSI representations.
    A multi-scale frequency information enhancement module then retains dominant spectral components at multiple scales to strengthen key features and suppress noise.
    Next, a CNN-KAN feature extraction module captures local correlations via cascaded convolutions and models long-range dependencies via Chebyshev KAN layers.
    Finally, a dual-domain fusion module adaptively integrates features from both branches to produce the prediction.
    Experiments on 3GPP-compliant QuaDRiGa datasets demonstrate that \OurMethod outperforms RNN, LSTM, GRU, CNN, and Transformer baselines in normalized mean square error (NMSE), spectral efficiency (SE), and bit error rate (BER) across various velocities and signal-to-noise ratios.
    Ablation studies further confirm the effectiveness of each proposed module.
\end{abstract}

\begin{IEEEkeywords}
    Channel prediction, CNN-KAN, multi-scale frequency enhancement, dual-domain, MIMO-OFDM.
\end{IEEEkeywords}

\section{Introduction}

Accurate channel state information (CSI) prediction is essential for improving the reliability, efficiency, and spectral capacity of massive MIMO-OFDM systems. In high-mobility environments, rapid channel fluctuations induced by Doppler shifts and multipath propagation cause the estimated CSI to become outdated, a phenomenon known as channel aging.
In time-division duplex (TDD) systems, downlink CSI can be inferred from uplink CSI via channel reciprocity, making timely and accurate CSI prediction critical for efficient resource allocation and reduced feedback overhead \cite{jiang-2019-neural, park-2025-EndtoEnd}.

Channel prediction methods generally fall into two categories: model-based and deep learning-based approaches.
Model-based methods, such as autoregressive models and Kalman filtering, depend on predefined channel statistical assumptions, limiting their adaptability to complex and dynamic wireless environments \cite{yin-2020-addressing, kim-2021-MassiveMIMO}.
Deep learning methods, in contrast, can learn channel dynamics directly from data and have achieved promising results~\cite{gao-2021-model, Song-2026-DSTND}.
However, accurate CSI prediction under high mobility requires jointly capturing two distinct types of temporal dependencies: \textit{short-term local variations} arising from rapid Doppler-induced fluctuations between adjacent time steps, and \textit{long-range nonlinear dependencies} that reflect the overall temporal evolution across the entire observation window. Existing deep learning approaches struggle to model both simultaneously.

Specifically, RNN-based models (including LSTM and GRU) process sequences in a step-by-step manner, making them effective at capturing short-term temporal correlations but susceptible to gradient vanishing when modeling long-range dependencies \cite{jiang-2020-recurrent, jiang-2020-deep, Cho-2014-LearningPhrase, hu-2024-ChannelPrediction}.
CNNs efficiently extract local spatial-frequency features within each time step through convolutional kernels, yet their fixed receptive fields inherently restrict the ability to capture temporal dependencies beyond a local neighborhood \cite{Safari-2020-DeepUL2DL, chen-2024-FrequencyDomain}.
Transformer-based models leverage self-attention to model global dependencies~\cite{Song-2026-CTPNet, zhou-2024-TransformerNetwork, mourya-2024-SpectralTemporal}, but their quadratic complexity and reliance on linear projections limit their capacity to fit the highly nonlinear temporal evolution of wireless channels \cite{Jiang-2022-AccurateChannel}.
In summary, none of the above architectures can simultaneously achieve efficient local feature extraction and flexible long-range nonlinear dependency modeling, which motivates the exploration of a new hybrid architecture.

To this end, we propose \OurMethod, a novel channel prediction model that combines CNNs and Kolmogorov-Arnold Networks (KANs)~\cite{liu-2025-kan} to jointly capture short-term local features and long-range nonlinear temporal dependencies.
The key insight is that CNNs and KANs are naturally complementary: CNNs excel at extracting \textit{intra-time-step local spatial-frequency correlations} through convolutional operations, while KANs, equipped with learnable Chebyshev polynomial activations on edges, can directly fit the \textit{inter-time-step nonlinear temporal evolution} of the entire CSI sequence in a holistic manner. Unlike conventional fully connected layers that rely on fixed activation functions, KANs place learnable univariate functions on network edges, offering superior expressiveness for modeling complex nonlinear mappings with fewer parameters.

Specifically, \OurMethod comprises four key components.
First, the Dual-Domain Expansion module (DDE) generates complementary frequency-domain and delay-domain CSI representations via IDFT~\cite{Liu-2024-LLM4CPAdaptingLarge}, providing richer input features for subsequent processing through two parallel branches.
Second, the Multi-Scale Frequency Information Enhancement module (MSFE) applies FFT along the temporal dimension and retains dominant frequency components at multiple scales, effectively enhancing key spectral patterns while suppressing noise and redundant fluctuations.
Third, the Dual-Domain CNN-KAN Feature Extraction module (DFE) employs cascaded convolutional layers to capture local spatial-temporal correlations within each time step, followed by Chebyshev KAN layers that perform holistic nonlinear mapping on the entire feature matrix to model long-range temporal dependencies across the sequence.
Finally, the Dual-Domain Fusion module (DDF) adaptively integrates features from both branches to produce the final prediction.

We conduct extensive experiments on channel datasets generated by the QuaDRiGa channel simulator~\cite{jaeckel-2014-quadriga} following 3GPP specifications.
The results demonstrate that \OurMethod consistently outperforms baseline methods, including RNN, LSTM, GRU, CNN, and Transformer-based models, in terms of normalized mean square error (NMSE), spectral efficiency (SE), and bit error rate (BER) across various velocities and signal-to-noise ratios (SNRs).
The main contributions of this paper are summarized as follows:
\begin{itemize}
    \item We propose a hybrid CNN-KAN architecture for CSI prediction that explicitly decouples the modeling of short-term and long-term dependencies: CNNs capture intra-time-step local spatial-frequency correlations, while Chebyshev polynomial-based KANs model inter-time-step long-range nonlinear temporal evolution. To the best of our knowledge, this is the first application of KAN to channel prediction.
    \item We design a multi-scale frequency information enhancement module that selects dominant spectral components at multiple scales via FFT-based filtering, effectively strengthening key channel features and suppressing noise.
    \item We introduce a dual-domain processing and fusion framework that jointly exploits frequency-domain and delay-domain CSI representations, providing complementary information that enhances overall prediction robustness.
    \item Extensive experiments on 3GPP-compliant QuaDRiGa datasets validate the effectiveness of the proposed model under diverse mobility conditions and SNR levels, with ablation studies confirming the contribution of each module.
\end{itemize}



\section{System Model and Problem Formulation}

We consider a TDD MIMO-OFDM system where the uplink and downlink transmissions share the same frequency band but are separated in time.
Due to channel reciprocity, the downlink CSI can be inferred from the uplink CSI after calibration.
However, in high-mobility scenarios, the time interval between CSI estimation and data transmission causes the acquired CSI to become outdated, a phenomenon known as channel aging.
Therefore, accurate future CSI prediction is essential to compensate for this aging effect and maintain system performance.

\subsection{Channel Model}

Consider a MIMO-OFDM system where the base station (BS) is equipped with a uniform planar array (UPA) of $N_t = N_h \times N_v$ antennas, and the user equipment (UE) has multiple omnidirectional antennas.
The system operates over $K$ subcarriers.
The downlink channel vector at time $t$ and subcarrier frequency $f_k$ is modeled as a superposition of multipath components:
\begin{equation}
    \begin{aligned}
        \mathbf{h}(t, f_k) & = \sum_{n=1}^{N_c} \sum_{m=1}^{M_n} \beta_{n,m}                                                                         \\
                           & \quad e^{j 2 \pi \left( \nu_{n,m} t - f_k \tau_{n,m} \right)} e^{j \phi_{n,m}} \mathbf{a}(\theta_{n,m}, \varphi_{n,m}),
    \end{aligned}
\end{equation}
where $N_c$ is the number of scattering clusters, $M_n$ is the number of rays in the $n$-th cluster, and $\beta_{n,m}$, $\nu_{n,m}$, $\tau_{n,m}$, $\phi_{n,m}$ denote the complex gain, Doppler frequency, propagation delay, and initial phase of the $(n,m)$-th path, respectively.
The array response vector $\mathbf{a}(\theta, \varphi) \in \mathbb{C}^{N_t}$ for the UPA is given by the Kronecker product of horizontal and vertical steering vectors:
\begin{equation}
    \mathbf{a}(\theta, \varphi) = \mathbf{a}_h(\theta, \varphi) \otimes \mathbf{a}_v(\theta).
\end{equation}

The CSI matrix at time $t$ is obtained by stacking the channel vectors across all $K$ subcarriers:
\begin{equation}
    \mathbf{H}^t = \left[\mathbf{h}(t,f_1), \mathbf{h}(t,f_2), \ldots, \mathbf{h}(t,f_K)\right]^T \in \mathbb{C}^{K \times N_t}.
\end{equation}

Under TDD channel reciprocity, the downlink CSI can be obtained from the uplink CSI through calibration matrices:
\begin{equation}
    \mathbf{H}_d^t = \mathbf{C}_{\text{BS}} \mathbf{H}_u^t \mathbf{C}_{\text{UE}},
\end{equation}
where $\mathbf{C}_{\text{BS}}$ and $\mathbf{C}_{\text{UE}}$ are the calibration matrices at the BS and UE sides, respectively.

\subsection{Channel Prediction Problem Formulation}

Given the historical CSI sequence over $T$ consecutive time steps:
\begin{equation}
    \mathbf{X}^{s} = \left\{\mathbf{H}^{s-T+1}, \mathbf{H}^{s-T+2}, \ldots, \mathbf{H}^{s}\right\},
\end{equation}
the objective is to predict the future CSI over the next $L$ time steps:
\begin{equation}
    \hat{\mathbf{Y}}^{s} = \left\{\hat{\mathbf{H}}^{s+1}, \hat{\mathbf{H}}^{s+2}, \ldots, \hat{\mathbf{H}}^{s+L}\right\}.
\end{equation}

This task can be formulated as learning a parameterized mapping $f_{\Omega}$:
\begin{equation}
    (\hat{\mathbf{H}}_d^{s+1}, \ldots, \hat{\mathbf{H}}_d^{s+L}) = f_{\Omega}(\mathbf{X}^s),
\end{equation}
where $\Omega$ denotes the set of all trainable parameters. The model is optimized by minimizing the normalized mean square error (NMSE) between the predicted and ground-truth CSI:
\begin{equation}
    \min_{\Omega} \ \mathrm{NMSE} = \mathbb{E} \left[\frac{\sum_{i=1}^{L} \|\hat{\mathbf{H}}_d^{s+i} - \mathbf{H}_d^{s+i}\|_F^2}{\sum_{i=1}^{L} \|\mathbf{H}_d^{s+i}\|_F^2} \right].
\end{equation}

The core challenge lies in effectively capturing both short-term local variations and long-range nonlinear temporal dependencies from the historical CSI sequence $\mathbf{X}^s$, which motivates the hybrid CNN-KAN architecture proposed in the next section.

\begin{figure*}[t]
    \centering
    \includegraphics[width=\textwidth]{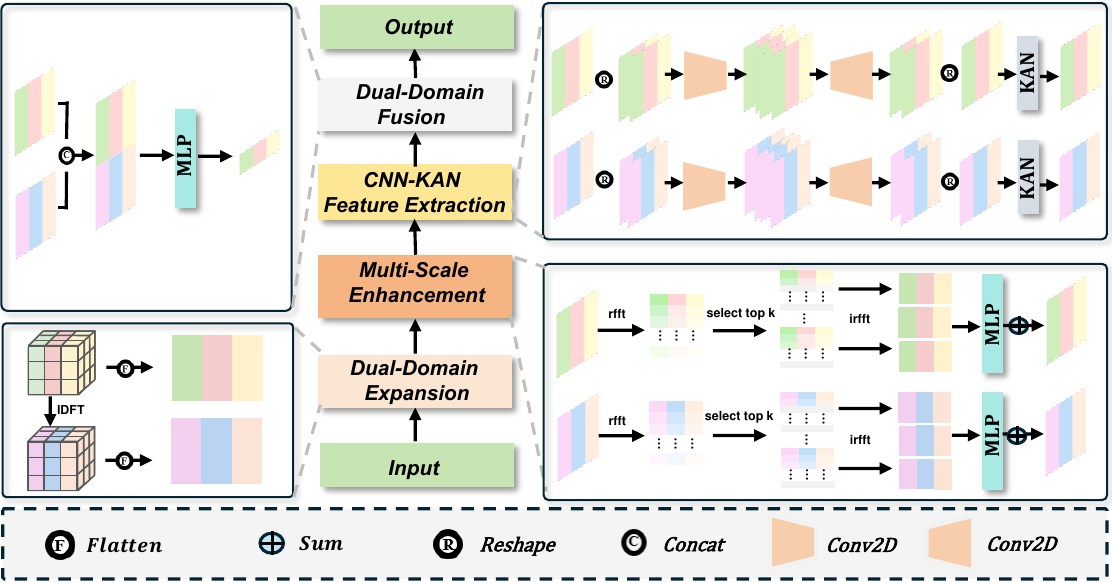}
    \caption{Overall architecture of the proposed \OurMethod. The model processes CSI through dual-domain expansion, multi-scale frequency enhancement, and parallel CNN-KAN branches that capture local spatio-temporal correlations and long-range nonlinear temporal dependencies, respectively.}
    \label{fig:model}
\end{figure*}

\section{Proposed Model}
This section presents \OurMethod, a CNN-KAN-based channel prediction model for TDD MIMO-OFDM systems. The model comprises four components: the Dual-Domain Expansion Module, the Multi-Scale Frequency Information Enhancement Module, the Dual-Domain CNN-KAN Feature Extraction Module, and the Dual-Domain Fusion Module. Each component addresses a specific aspect of CSI modeling, collectively enabling the capture of both short-term local correlations and long-range nonlinear temporal dependencies.

\subsection{\textbf{Dual-Domain Expansion Module}}
The Dual-Domain Expansion Module generates two complementary CSI representations in the frequency domain and the delay domain. By providing information from both domains, this module enables the subsequent branches to exploit distinct structural properties of the channel, which is essential for capturing temporal dynamics under high-mobility conditions.

For a given real-valued CSI input $\mathbf{X} \in \mathbb{R}^{T \times K \times (N_t \cdot N_r)}$, we define the feature dimension $C = K \cdot N_t \cdot N_r$. The module generates two representations as follows.

\textbf{Frequency Domain Representation (CFR).} The CSI is reshaped into a two-dimensional temporal representation while preserving subcarrier-domain correlations:
\begin{equation}\label{eq:cfr}
    \mathbf{X}_{\text{F}} = \text{Flatten}(\mathbf{X}) \in \mathbb{R}^{T \times C}.
\end{equation}

\textbf{Delay Domain Representation (CIR).} The CSI is transformed from the frequency domain to the delay domain via the inverse discrete Fourier transform (IDFT), revealing sparse multipath components:
\begin{equation}\label{eq:idft}
    \mathbf{G}^t = \text{IDFT}_K(\mathbf{H}^t), \quad t = 1, 2, \ldots, T.
\end{equation}
The resulting CIR is flattened by separating the real and imaginary parts:
\begin{equation}\label{eq:cir_flatten}
    \mathbf{X}_{\text{D}} = \text{Flatten}\left(\left[\Re\{\mathbf{G}\}, \Im\{\mathbf{G}\}\right]\right) \in \mathbb{R}^{T \times C}.
\end{equation}

The complementary features from both branches are concatenated along the channel dimension to form the dual-domain feature vector:
\begin{equation}\label{eq:dual_concat}
    \mathbf{X}_{\text{dual}} = \text{Concat}(\mathbf{X}_{\text{F}}, \mathbf{X}_{\text{D}}) \in \mathbb{R}^{T \times 2C}.
\end{equation}

\subsection{\textbf{Multi-Scale Frequency Information Enhancement Module}}
The Multi-Scale Frequency Information Enhancement Module strengthens the most informative spectral components of the CSI sequence at multiple scales, thereby suppressing irrelevant fluctuations and noise while preserving key spectral patterns.

Taking the dual-domain features as input, we first apply the \textbf{real-valued fast Fourier transform (rFFT)} along the temporal dimension to convert the CSI sequence into the spectral domain:
\begin{equation}\label{eq:rfft}
    \mathbf{S} = \text{rFFT}(\mathbf{Z}_0) \in \mathbb{C}^{\left(\left\lfloor \frac{T}{2} \right\rfloor + 1\right) \times C},
\end{equation}
where $\mathbf{Z}_0$ denotes the input feature of this module. For each scale $q$, the top-$r_q$ dominant frequency components are selected to retain critical spectral information:
\begin{equation}\label{eq:topk}
    \mathcal{I}_q = \text{TopK}(|\mathbf{S}|, r_q), \quad q = 1, 2, \ldots, k.
\end{equation}
A binary mask is then constructed to filter out trivial frequency components:
\begin{equation}\label{eq:mask}
    \mathbf{M}_q(\omega, c) =
    \begin{cases}
        1, & \omega \in \mathcal{I}_q(c), \\
        0, & \text{otherwise}.
    \end{cases}
\end{equation}
The denoised spectral feature at scale $q$ is obtained via element-wise multiplication of the mask and the original spectrum:
\begin{equation}\label{eq:denoise_spec}
    \tilde{\mathbf{S}}_q = \mathbf{M}_q \odot \mathbf{S}.
\end{equation}
The inverse real-valued FFT (irFFT) is then applied to reconstruct the enhanced time-domain features from the filtered spectrum:
\begin{equation}\label{eq:irfft}
    \tilde{\mathbf{Z}}_q = \text{irFFT}(\tilde{\mathbf{S}}_q) \in \mathbb{R}^{T \times C}.
\end{equation}
Finally, the multi-scale enhanced features are aggregated through a learnable dense layer to produce the output of this module:
\begin{equation}\label{eq:ms_feature}
    \mathbf{Z}_{\text{ms}} = \sum_{q=1}^{k} \mathbf{U}_q \in \mathbb{R}^{T \times C}.
\end{equation}
where $\mathbf{U}_q$ denotes the learnable weight matrix for the $q$-th scale feature.

\subsection{\textbf{Dual-Domain CNN-KAN Feature Extraction Module}}
The Dual-Domain CNN-KAN Feature Extraction Module operates through two parallel branches for the CFR and CIR domains. Cascaded convolutional layers capture \textit{intra-time-step local spatial-frequency correlations}, while the Chebyshev polynomial-based KAN models \textit{inter-time-step long-range temporal dependencies} through holistic matrix-level nonlinear mapping.

For each domain $d \in \{\text{F}, \text{D}\}$, the multi-scale feature $\mathbf{Z}_{\text{ms},d}$ is reshaped to match the input dimension of convolutional layers:
\begin{equation}\label{eq:reshape_cnn}
    \mathbf{Z}_d^0 = \text{Reshape}(\mathbf{Z}_{\text{ms},d}) \in \mathbb{R}^{T \times C \times 2}.
\end{equation}
Cascaded convolutional layers with nonlinear activation $\sigma(\cdot)$ extract local spatio-temporal correlations within each time step:
\begin{equation}\label{eq:cnn_forward}
    \mathbf{Q}_d = \sigma\big(\text{Conv}_L \big(\cdots \sigma\big(\text{Conv}_1(\mathbf{Z}_d^0)\big) \cdots\big)\big).
\end{equation}
The convolutional output is then reshaped into a two-dimensional matrix for holistic temporal dependency modeling:
\begin{equation}\label{eq:reshape_kan}
    \bar{\mathbf{Q}}_d = \text{Reshape}(\mathbf{Q}_d) \in \mathbb{R}^{T \times C}.
\end{equation}

\subsubsection{Chebyshev KAN with Holistic Matrix Representation}
Unlike conventional linear layers, the Chebyshev KAN employs learnable Chebyshev polynomial basis functions to fit the nonlinear temporal evolution of CSI. The first-kind Chebyshev polynomials are defined by the following recurrence:
\begin{equation}\label{eq:cheby_rec}
    \begin{cases}
        T_0(x) = 1, \\
        T_1(x) = x, \\
        T_{m+1}(x) = 2xT_m(x) - T_{m-1}(x), \quad m\ge 1.
    \end{cases}
\end{equation}
Rather than decomposing individual time series, the KAN performs holistic nonlinear mapping directly on the entire feature matrix $\bar{\mathbf{Q}}_d \in \mathbb{R}^{T \times C}$, where each row corresponds to a time step and each column represents a feature dimension. The input matrix is first normalized to the interval $[-1,1]$ via $\tanh(\cdot)$ to match the domain of Chebyshev polynomials:
\begin{equation}
    \hat{\mathbf{Q}}_d = \tanh\big(\bar{\mathbf{Q}}_d\big).
\end{equation}
The matrix-form Chebyshev mapping operator is defined as:
\begin{equation}\label{eq:matrix_phi}
    \mathbf{\Phi}(\hat{\mathbf{Q}}_d) = \sum_{m=0}^{M} \mathbf{W}_m \odot T_m(\hat{\mathbf{Q}}_d),
\end{equation}
where $\mathbf{W}_m$ denotes the learnable coefficient matrix, $M$ is the maximum polynomial order, and $\odot$ denotes element-wise multiplication. By applying the holistic Chebyshev KAN mapping, the final feature matrix encoding long-range temporal dependencies is obtained as:
\begin{equation}
    \mathbf{F}_d = \mathbf{\Phi}(\hat{\mathbf{Q}}_d) \in \mathbb{R}^{T \times C}.
\end{equation}

\subsection{\textbf{Dual-Domain Fusion}}
The Dual-Domain Fusion module concatenates the features extracted from the CFR and CIR branches to integrate complementary dual-domain information:
\begin{equation}\label{eq:fusion_concat}
    \mathbf{F}_{\text{dual}} = \mathbf{\text{Concat}}(\mathbf{F}_\text{F}, \mathbf{F}_\text{D}) \in \mathbb{R}^{T \times 2C}.
\end{equation}
A dense fusion layer maps the concatenated features to the prediction dimension:
\begin{equation}\label{eq:dense_fusion}
    \mathbf{O} = \sigma(\mathbf{F}_{\text{dual}} \mathbf{W}_f + \mathbf{b}_f) \in \mathbb{R}^{P \times C},
\end{equation}
where $\mathbf{W}_f$ and $\mathbf{b}_f$ are trainable weight and bias, and $P$ denotes the length of the predicted CSI sequence. Finally, the output is reshaped and inversely transformed to recover the complex-valued CSI:
\begin{equation}\label{eq:final_pred}
    \hat{\mathbf{Y}} = \mathcal{R}^{-1}(\text{Reshape}(\mathbf{O})) \in \mathbb{C}^{P \times K \times (N_t \cdot N_r)}.
\end{equation}

\section{EXPERIMENTAL SETUP AND RESULTS}

This section evaluates the proposed \OurMethod through comprehensive experiments. We first describe the experimental configuration, then assess the prediction performance under different mobility conditions and SNR levels, and finally conduct ablation studies to verify the contribution of each module.

\subsection{Experimental Setup}
Following the configuration of LLM4CP \cite{Liu-2024-LLM4CPAdaptingLarge} for fair comparison, we adopt a MIMO-OFDM system with a UPA-equipped base station.
Time-varying channel datasets are generated by QuaDRiGa \cite{jaeckel-2014-quadriga} under 3GPP specifications \cite{3GPP-2018-TR38901}.
The historical window is $T=16$ and the prediction horizon is $L=4$.
Each mobility condition (10--100 km/h) contains 8000 training, 1000 validation, and 1000 test samples.
The model is trained using the Adam optimizer with learning rate decay and NMSE as the loss function.
Performance is evaluated by NMSE, spectral efficiency (SE) \cite{Liu-2024-LLM4CPAdaptingLarge}, and bit error rate (BER) \cite{Liu-2024-LLM4CPAdaptingLarge}.
We compare against five baselines: RNN, LSTM, GRU, CNN, and Transformer.

\subsection{Performance Under Different Mobilities}

\begin{figure}[t]
    \centering
    \includegraphics[width=\columnwidth]{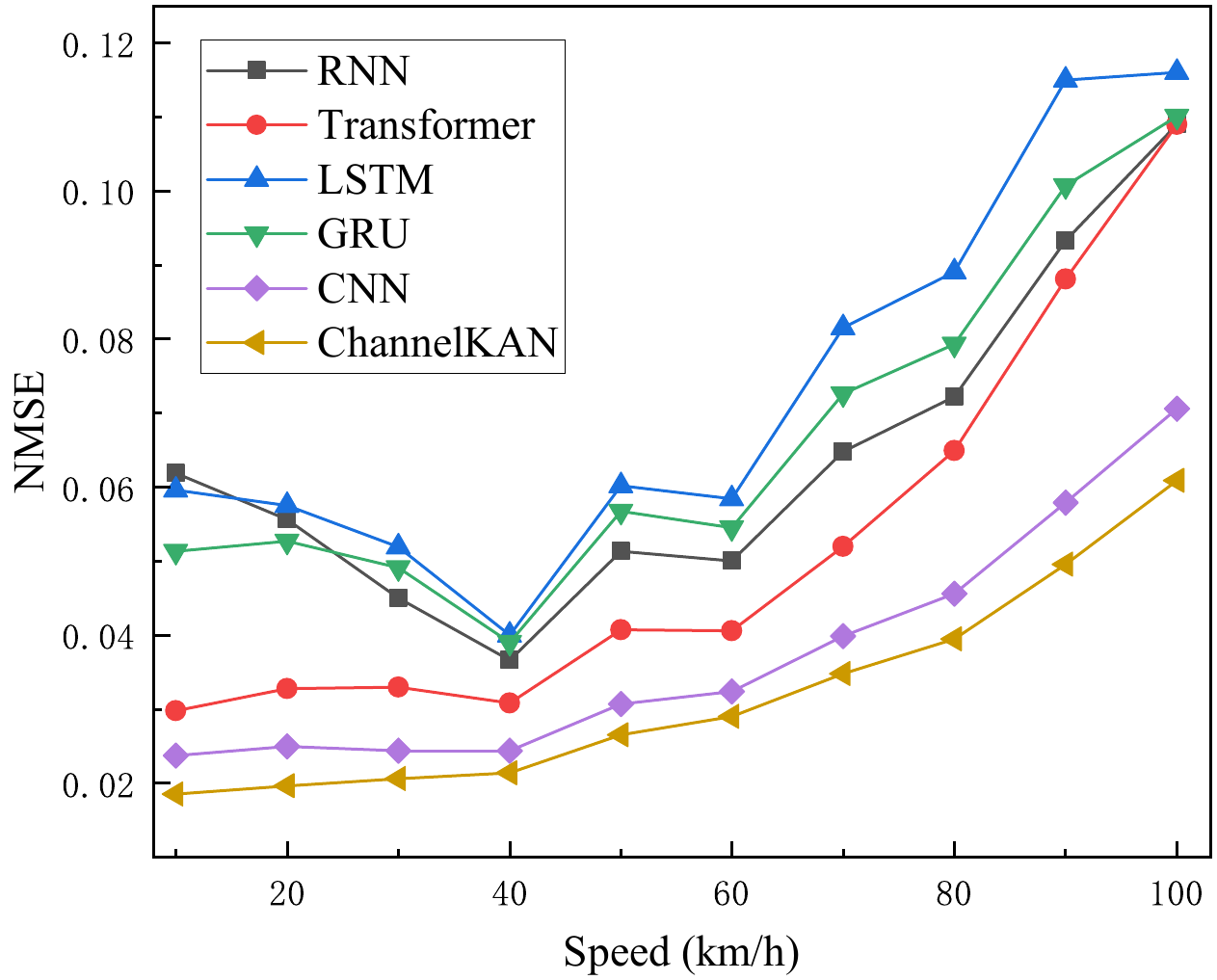} 
    \caption{NMSE performance comparison of different prediction methods under user velocities ranging from 10 to 100 km/h.}
    \label{fig:speed}
\end{figure}

Fig.~\ref{fig:speed} compares the NMSE of all methods under velocities from 10 to 100 km/h. As velocity increases, the Doppler effect intensifies and channel coherence time shortens, making prediction increasingly challenging.
As shown in the figure, \OurMethod achieves the lowest NMSE across all velocities, with a particularly significant advantage above 60 km/h (NMSE $\approx$ 0.0609 at 100 km/h).
Among the baselines, RNN-based models degrade rapidly at high velocities due to gradient vanishing, CNN performs reasonably at medium velocities but is limited by its fixed receptive field, and the NMSE of Transformer rises sharply at high velocities despite its global attention mechanism.
These results confirm that the CNN-KAN hybrid architecture effectively balances short-term and long-range dependency modeling under diverse mobility conditions.

\subsection{Performance Under Different SNR}

\begin{figure}[t]
    \centering
    \includegraphics[width=\columnwidth]{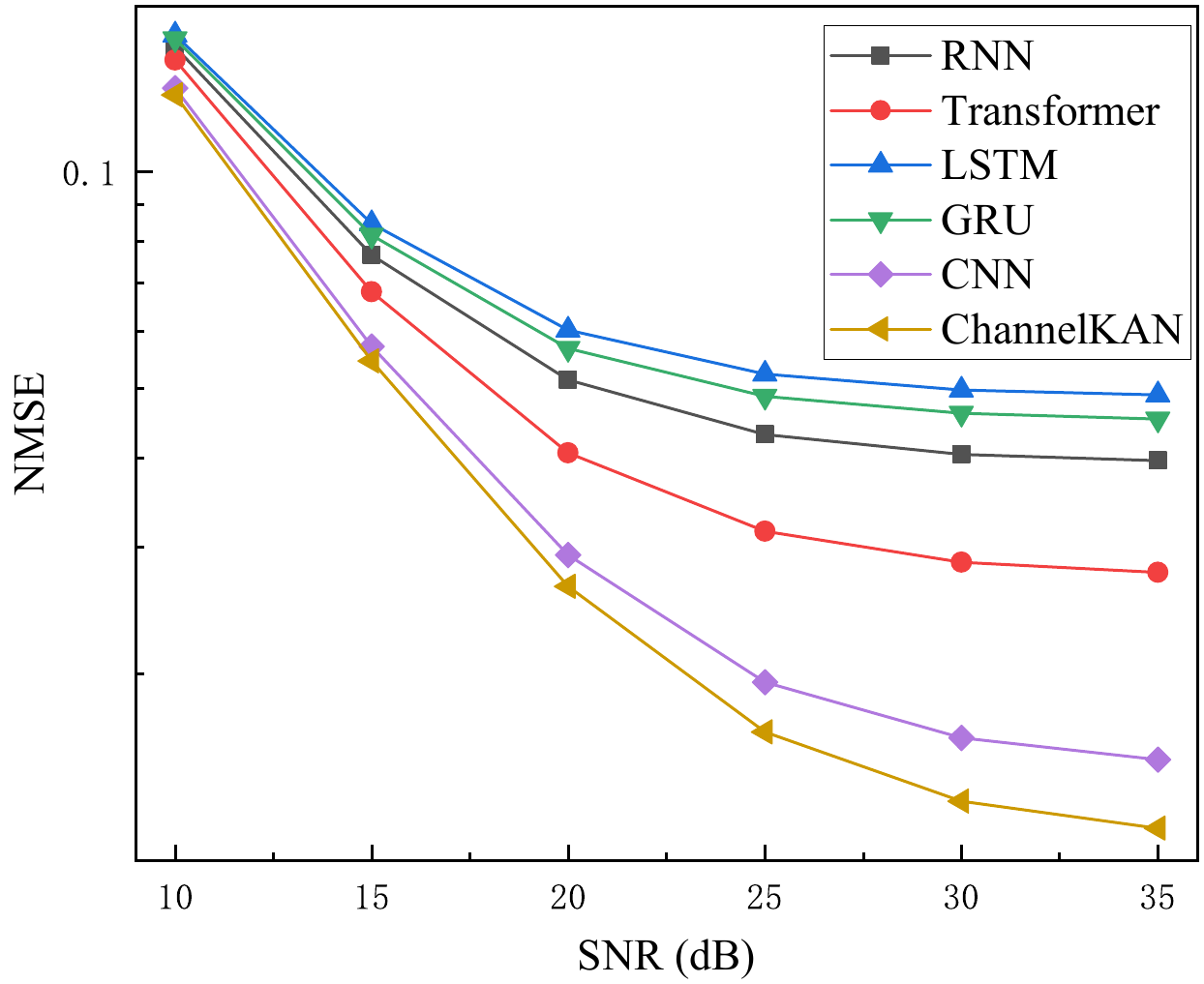}
    \caption{NMSE performance comparison of different prediction methods under varying SNR levels with noisy historical CSI.}
    \label{fig:noise}
\end{figure}

Fig.~\ref{fig:noise} presents the NMSE under SNR ranging from 0 to 30 dB, where noise is added to the historical CSI. \OurMethod consistently outperforms all baselines across the entire SNR range, with a particularly notable advantage at low SNR levels. This robustness is attributed to the multi-scale frequency enhancement module, which effectively suppresses noise while preserving dominant channel features.

\subsection{Spectral Efficiency, BER, and Parameter Size Analysis}

\begin{table}[t]
    \centering
    \tt

    \caption{Spectral efficiency (SE), bit error rate (BER), and parameter size comparison of different prediction methods.}
    \label{tab:spectral_efficiency}
    \setlength{\tabcolsep}{3pt}
    \begin{adjustbox}{width=1\linewidth}
        \begin{tabular}{lllllll}
            \toprule
            Metric & RNN     & Transformer & LSTM    & GRU     & CNN     & \OurMethod   \\ \midrule
            SE (bps/Hz)     & 6.328   & 6.345       & 6.309   & 6.328   & 6.374   & \textbf{6.414}   \\
            BER             & 0.00849 & 0.00845     & 0.00860 & 0.00847 & 0.00791 & \textbf{0.00771} \\
            Params          & 3.8MB   & 65MB        & 4.2MB   & \textbf{3.2MB} & 12MB & 24.5MB \\ \bottomrule
        \end{tabular}
    \end{adjustbox}
\end{table}

Table~\ref{tab:spectral_efficiency} summarizes the SE and BER performance. \OurMethod achieves the highest SE of 6.414 bps/Hz and the lowest BER of 0.00771, outperforming all baselines on both metrics. These results indicate that more accurate CSI prediction directly translates into improved communication capacity and transmission reliability. Despite the superior performance, \OurMethod achieves this with a relatively moderate parameter size of 24.5MB, which is competitive compared to other methods such as Transformer (65MB) and CNN (12MB).

\subsection{Ablation Study}

\begin{table}[t]
    \centering
    \tt

    \caption{Ablation study results showing the impact of each module on NMSE, SE, and BER.}
    \label{tab:ablations}
    \setlength{\tabcolsep}{3pt}

    \begin{adjustbox}{width=1\linewidth}
        \begin{tabular}{llllll}
            \toprule
            Metric & w/o Multi-Scale & w/o CNN-KAN & w/o Dual-Domain & w/o KAN & \OurMethod   \\
            \midrule
            NMSE            & 0.0274                   & 0.0628               & 0.0318                   & 0.0282           & \textbf{0.0265}  \\
            SE(bps/Hz)      & 6.401                    & 6.237                & 6.385                    & 4.400            & \textbf{6.414}   \\
            BER             & 0.0079                   & 0.0089               & 0.00799                  & 0.0080           & \textbf{0.00771} \\
            \bottomrule
        \end{tabular}
    \end{adjustbox}
\end{table}

To assess the contribution of each module, we evaluate four ablation variants, with results shown in Table~\ref{tab:ablations}.
The CNN-KAN module contributes the most, as its removal causes the largest NMSE increase (from 0.0265 to 0.0628), confirming its central role in jointly modeling local and long-range dependencies.
Removing dual-domain processing raises NMSE to 0.0318, demonstrating the value of complementary frequency-domain and delay-domain features.
Without multi-scale enhancement, NMSE increases to 0.0274, validating its effectiveness in strengthening dominant spectral patterns.
Removing KAN alone raises NMSE to 0.0282 and sharply degrades SE to 4.400 bps/Hz, highlighting the critical contribution of KAN to nonlinear temporal modeling.

\subsection{Prediction Visualization}

\begin{figure}[t]
    \centering

    \begin{subfigure}[b]{0.48\columnwidth}
        \centering
        \includegraphics[width=\linewidth]{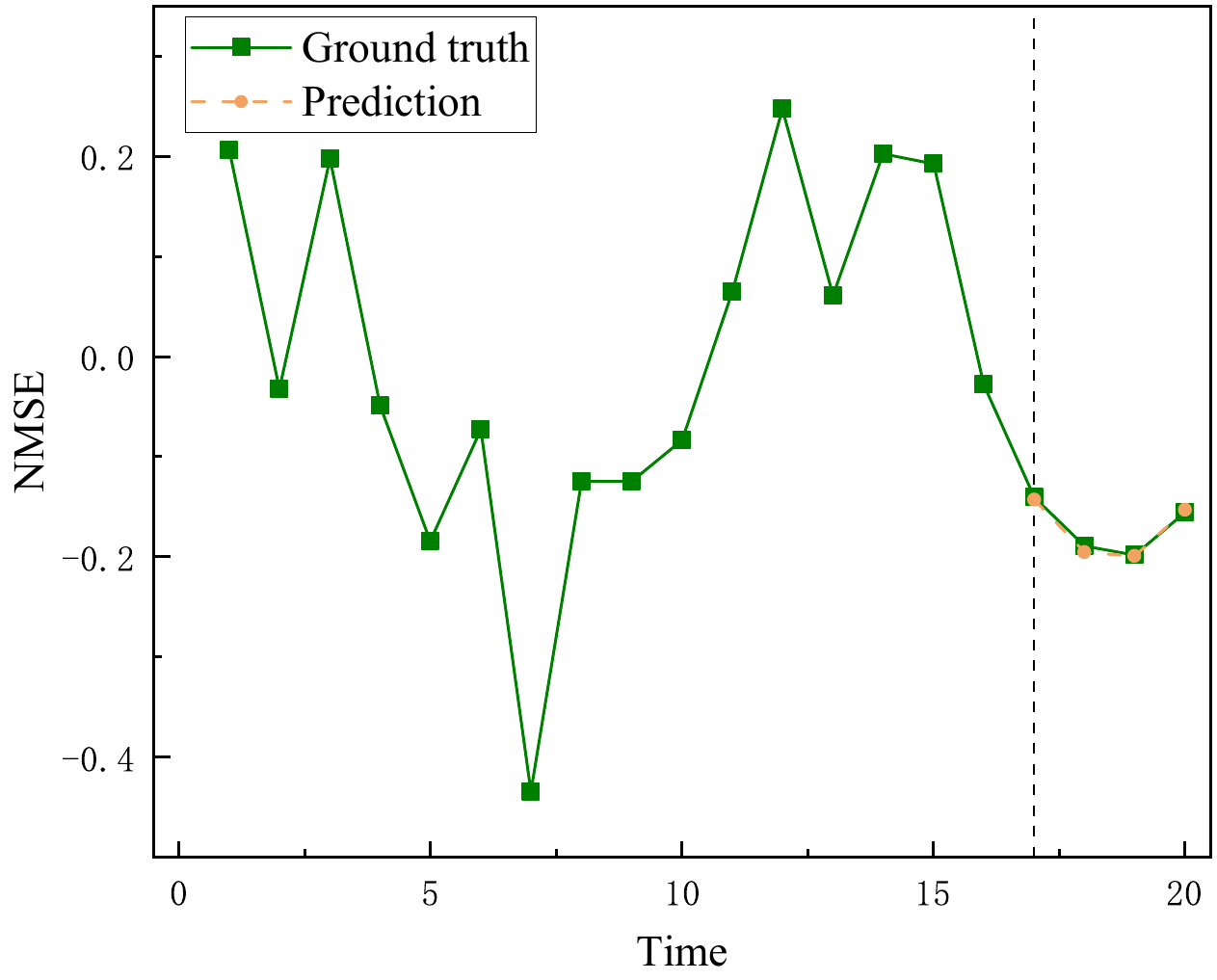}
        \caption{\scriptsize Speed = 80 km/h, SNR = 10 dB}
        \label{fig:speed80noise10}
    \end{subfigure}
    \hfill
    \begin{subfigure}[b]{0.48\columnwidth}
        \centering
        \includegraphics[width=\linewidth]{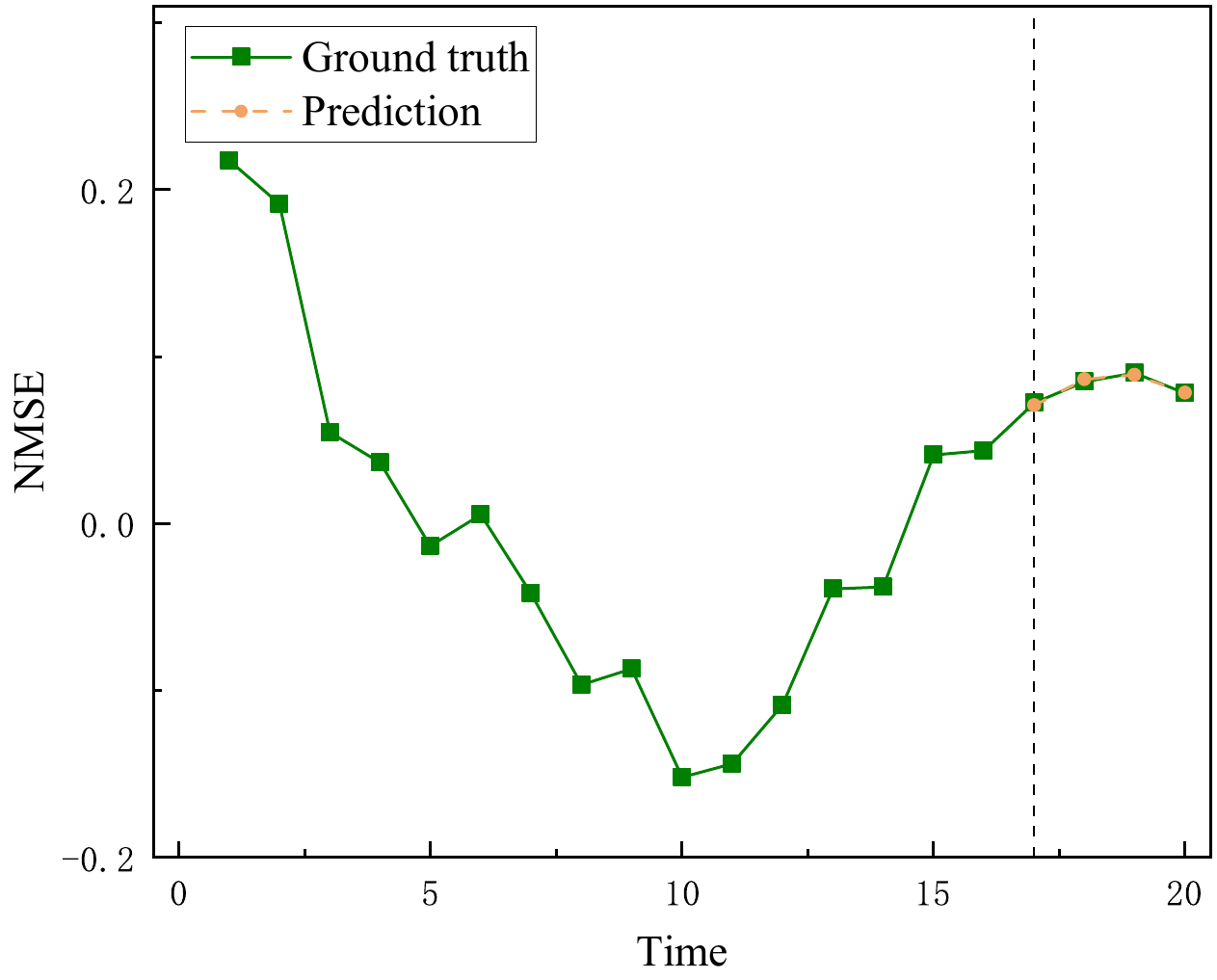}
        \caption{\scriptsize Speed = 80 km/h, SNR = 20 dB}
        \label{fig:speed80noise20}
    \end{subfigure}

    \vspace{0.2cm}

    \begin{subfigure}[b]{0.48\columnwidth}
        \centering
        \includegraphics[width=\linewidth]{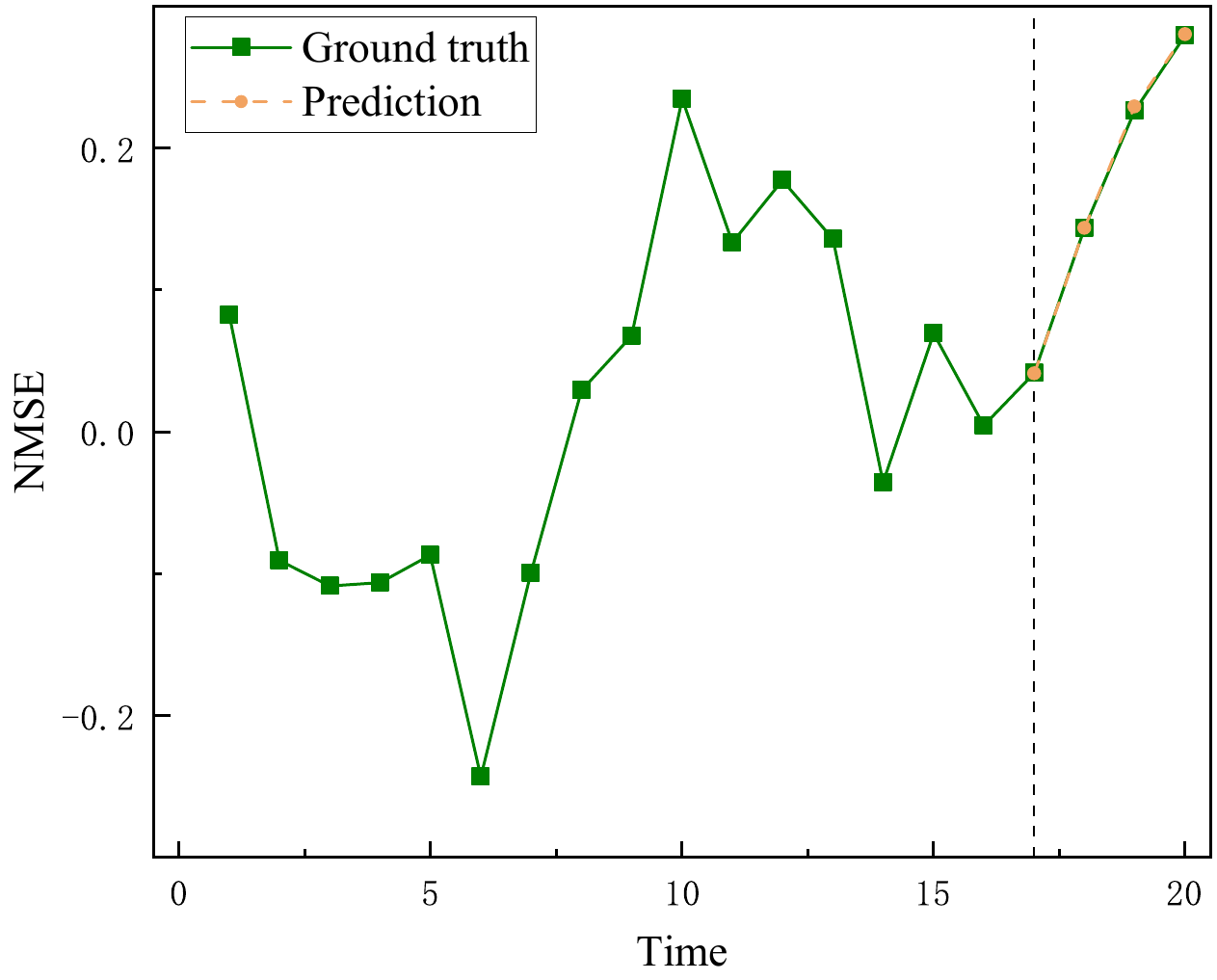}
        \caption{\scriptsize Speed = 90 km/h, SNR = 20 dB}
        \label{fig:speed90noise20}
    \end{subfigure}
    \hfill
    \begin{subfigure}[b]{0.48\columnwidth}
        \centering
        \includegraphics[width=\linewidth]{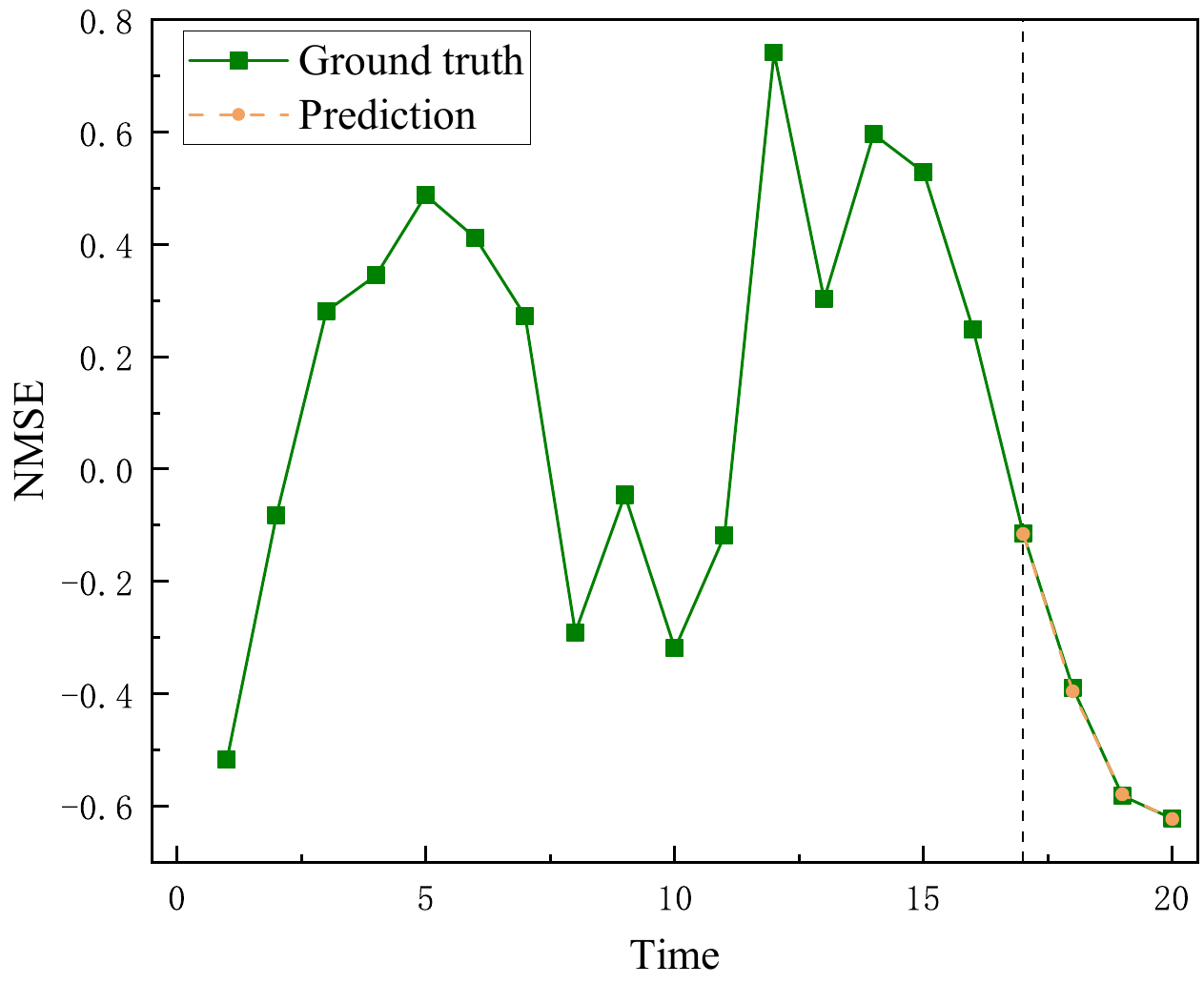}
        \caption{\scriptsize Speed = 100 km/h, SNR = 10 dB}
        \label{fig:speed100noise10}
    \end{subfigure}

    \caption{Visualization of predicted versus ground-truth CSI under different speed and SNR conditions.}
    \label{fig:speed_noise_prediction}
\end{figure}

Fig.~\ref{fig:speed_noise_prediction} visualizes the predicted CSI against the ground truth under four representative speed-SNR combinations. The predicted curves closely track the ground truth across all conditions, demonstrating the ability of the model to capture channel dynamics even under high mobility and low SNR.

\section{CONCLUSION}

This paper proposes \OurMethod, a hybrid CNN-KAN channel prediction model for high-mobility MIMO-OFDM systems. By decoupling short-term local feature extraction (via CNN) from long-range nonlinear temporal modeling (via Chebyshev KAN), and integrating multi-scale frequency enhancement with dual-domain processing, \OurMethod achieves the best NMSE, SE, and BER across velocities from 10 to 100 km/h and various SNR levels on 3GPP-compliant QuaDRiGa datasets.
Future work will focus on reducing computational cost for real-time deployment and validating generalization on measured channel data.

\balance
\bibliographystyle{IEEEtran}
\bibliography{ref}

@article{jiang-2019-neural,
  author  = {Jiang, Wei and Schotten, Hans Dieter},
  doi     = {10.1109/ACCESS.2019.2937588},
  journal = {IEEE Access},
  number  = {},
  pages   = {118112-118124},
  title   = {Neural Network-Based Fading Channel Prediction: A Comprehensive Overview},
  volume  = {7},
  year    = {2019}
}

@article{yin-2020-addressing,
  author  = {Yin, Haifan and Wang, Haiquan and Liu, Yingzhuang and Gesbert, David},
  doi     = {10.1109/JSAC.2020.3005473},
  journal = {IEEE Journal on Selected Areas in Communications},
  number  = {12},
  pages   = {2903-2917},
  title   = {Addressing the Curse of Mobility in Massive MIMO With Prony-Based Angular-Delay Domain Channel Predictions},
  volume  = {38},
  year    = {2020}
}

@inproceedings{jiang-2020-recurrent,
  author    = {Jiang, Wei and Schotten, Hans Dieter},
  booktitle = {2020 IEEE 91st Vehicular Technology Conference (VTC2020-Spring)},
  doi       = {10.1109/VTC2020-Spring48590.2020.9128426},
  number    = {},
  pages     = {1-5},
  title     = {Recurrent Neural Networks with Long Short-Term Memory for Fading Channel Prediction},
  volume    = {},
  year      = {2020}
}

@article{jiang-2020-deep,
  author  = {Jiang, Wei and Schotten, Hans Dieter},
  doi     = {10.1109/OJCOMS.2020.2982513},
  journal = {IEEE Open Journal of the Communications Society},
  number  = {},
  pages   = {320-332},
  title   = {Deep Learning for Fading Channel Prediction},
  volume  = {1},
  year    = {2020}
}

@inproceedings{Cho-2014-LearningPhrase,
  author    = {Kyunghyun Cho and
               Bart van Merrienboer and
               {\c{C}}aglar G{\"{u}}l{\c{c}}ehre and
               Dzmitry Bahdanau and
               Fethi Bougares and
               Holger Schwenk and
               Yoshua Bengio},
  booktitle = {Proceedings of the 2014 Conference on Empirical Methods in Natural
               Language Processing (EMNLP)},
  doi       = {10.3115/V1/D14-1179},
  pages     = {1724--1734},
  publisher = {{ACL}},
  title     = {Learning Phrase Representations using {RNN} Encoder-Decoder for Statistical Machine Translation},
  url       = {https://doi.org/10.3115/v1/d14-1179},
  year      = {2014}
}

@article{Safari-2020-DeepUL2DL,
  author  = {Safari, Mohammad Sadegh and Pourahmadi, Vahid and Sodagari, Shabnam},
  doi     = {10.1109/OJVT.2019.2962631},
  journal = {IEEE Open Journal of Vehicular Technology},
  number  = {},
  pages   = {29-44},
  title   = {Deep UL2DL: Data-Driven Channel Knowledge Transfer From Uplink to Downlink},
  volume  = {1},
  year    = {2020}
}

@article{Jiang-2022-AccurateChannel,
  author  = {Jiang, Hao and Cui, Mingyao and Ng, Derrick Wing Kwan and Dai, Linglong},
  doi     = {10.1109/JSAC.2022.3191334},
  journal = {IEEE Journal on Selected Areas in Communications},
  number  = {9},
  pages   = {2717-2732},
  title   = {Accurate Channel Prediction Based on Transformer: Making Mobility Negligible},
  volume  = {40},
  year    = {2022}
}

@article{Liu-2024-LLM4CPAdaptingLarge,
  author  = {Boxun Liu and Xuanyu Liu and Shijian Gao and Xiang Cheng and Liuqing Yang},
  doi     = {10.23919/JCIN.2024.10582829},
  journal = {Journal of Communications and Information Networks},
  number  = {2},
  pages   = {113--125},
  title   = {{LLM4CP:} Adapting Large Language Models for Channel Prediction},
  volume  = {9},
  year    = {2024}
}

@article{jaeckel-2014-quadriga,
  author  = {Jaeckel, Stephan and Raschkowski, Leszek and Börner, Kai and Thiele, Lars},
  doi     = {10.1109/TAP.2014.2310220},
  journal = {IEEE Transactions on Antennas and Propagation},
  number  = {6},
  pages   = {3242-3256},
  title   = {QuaDRiGa: A 3-D Multi-Cell Channel Model With Time Evolution for Enabling Virtual Field Trials},
  volume  = {62},
  year    = {2014}
}

@techreport{3GPP-2018-TR38901,
  author      = {{3GPP Radio Access Network Working Group}},
  institution = {3rd Generation Partnership Project (3GPP)},
  number      = {TR 38.901},
  title       = {Study on Channel Model for Frequencies from 0.5 to 100 GHz (Release 15)},
  year        = {2018}
}

@inproceedings{Song-2026-CTPNet,
  address   = {Barcelona, Spain, May 4-8},
  author    = {Song, Zhangyao and Jiang, Nanqing and He, Miaohong and Zhao, Xiaoyu and Guo, Tao},
  booktitle = {International Conference on Acoustics, Speech and Signal Processing (ICASSP)},
  doi       = {10.1109/ICASSP55912.2026.11464481},
  pages     = {4821--4825},
  publisher = {{IEEE}},
  title     = {{Channel, Trend and Periodic-Wise Representation Learning for Multivariate Long-Term Time Series Forecasting}},
  year      = {2026}
}

@article{Song-2026-DSTND,
  author  = {Song, Zhangyao and Zhang, Xiang and Zhuang, Li and Guo, Tao and Zhao, Xiaoyu and Xu, Yinfei and Jin, Shi},
  doi     = {10.1109/TCCN.2026.3685404},
  journal = {IEEE Transactions on Cognitive Communications and Networking},
  number  = {},
  pages   = {7647-7661},
  title   = {Diffusion-Based Spatio-Temporal Channel Prediction via Non-Stationarity Decoupling},
  volume  = {12},
  year    = {2026}
}

@inproceedings{liu-2025-kan,
  author    = {Ziming Liu and Yixuan Wang and Sachin Vaidya and Fabian Ruehle and James Halverson and Marin Soljacic and Thomas Y. Hou and Max Tegmark},
  booktitle = {The Thirteenth International Conference on Learning Representations},
  title     = {{KAN}: Kolmogorov{\textendash}Arnold Networks},
  url       = {https://openreview.net/forum?id=Ozo7qJ5vZi},
  year      = {2025}
}

@article{kim-2021-MassiveMIMO,
  author  = {Kim, Hwanjin and Kim, Sucheol and Lee, Hyeongtaek and Jang, Chulhee and Choi, Yongyun and Choi, Junil},
  doi     = {10.1109/TCOMM.2020.3027882},
  journal = {IEEE Transactions on Communications},
  number  = {1},
  pages   = {518-528},
  title   = {Massive MIMO Channel Prediction: Kalman Filtering Vs. Machine Learning},
  volume  = {69},
  year    = {2021}
}

@article{gao-2021-model,
  author   = {Gao, Shijian and Cheng, Xiang and Fang, Luoyang and Yang, Liuqing},
  doi      = {10.1109/TWC.2021.3061212},
  journal  = {IEEE Transactions on Wireless Communications},
  number   = {7},
  pages    = {4646-4656},
  title    = {Model Enhanced Learning Based Detectors (Me-LeaD) for Wideband Multi-User 1-bit mmWave Communications},
  volume   = {20},
  year     = {2021}
}

@article{hu-2024-ChannelPrediction,
  author   = {Hu, Xin and Huo, Yiming and Dong, Xiaodai and Wu, Fei-Yun and Huang, Aiping},
  doi      = {10.1109/JIOT.2023.3296116},
  journal  = {IEEE Internet of Things Journal},
  number   = {2},
  pages    = {3250-3263},
  title    = {Channel Prediction Using Adaptive Bidirectional GRU for Underwater MIMO Communications},
  volume   = {11},
  year     = {2024}
}

@article{park-2025-EndtoEnd,
  author   = {Park, Juseong and Sohrabi, Foad and Ghosh, Amitava and Andrews, Jeffrey G.},
  doi      = {10.1109/TWC.2024.3516633},
  journal  = {IEEE Transactions on Wireless Communications},
  number   = {3},
  pages    = {2110-2125},
  title    = {End-to-End Deep Learning for TDD MIMO Systems in the 6G Upper Midbands},
  volume   = {24},
  year     = {2025}
}

@article{chen-2024-FrequencyDomain,
  author   = {Huang, Chen and Wang, Cheng-Xiang and Li, Zheao and Qian, Zhongyu and Li, Junling and Miao, Yang},
  doi      = {10.1109/TCOMM.2024.3376602},
  journal  = {IEEE Transactions on Communications},
  number   = {8},
  pages    = {4887-4902},
  title    = {A Frequency Domain Predictive Channel Model for 6G Wireless MIMO Communications Based on Deep Learning},
  volume   = {72},
  year     = {2024}
}

@article{mourya-2024-SpectralTemporal,
  author   = {Mourya, Sharan and Reddy, Pavan and Amuru, SaiDhiraj and Kuchi, Kiran Kumar},
  doi      = {10.1109/LWC.2024.3372148},
  journal  = {IEEE Wireless Communications Letters},
  number   = {5},
  pages    = {1399-1403},
  title    = {Spectral Temporal Graph Neural Network for Massive MIMO CSI Prediction},
  volume   = {13},
  year     = {2024}
}

@article{zhou-2024-TransformerNetwork,
  author   = {Zhou, Tao and Liu, Xiangping and Xiang, Zuowei and Zhang, Haitong and Ai, Bo and Liu, Liu and Jing, Xiaorong},
  doi      = {10.1109/TWC.2024.3379123},
  journal  = {IEEE Transactions on Wireless Communications},
  number   = {9},
  pages    = {11154-11167},
  title    = {Transformer Network Based Channel Prediction for CSI Feedback Enhancement in AI-Native Air Interface},
  volume   = {23},
  year     = {2024}
}


\end{document}